\newif\iffigs\figsfalse
  \figstrue
\iffigs \input epsf \else  \fi

\def\be{\begin{equation}}
\def\ee{\end{equation}}
\def\ba{\begin{eqnarray}}
\def\ea{\end{eqnarray}}
\def\bC{{\bf C}}
\def\bZ{{\bf Z}}
\def\t{{}^t\!}
\def\sptn{$Sp(2n-2,\bZ)$}
\def\uon{$U(1)$}

\def\sun{$SU(n)$}

\def\seff{S_{\rm eff}}

\documentstyle[preprint,aps,prl,floats]{revtex}
\begin{document}
\newlength{\pubnumber} \settowidth{\pubnumber}{IASSNS-HEP-94/999}
\preprint{\parbox{\pubnumber} {\begin{flushleft}
IASSNS-HEP-94/94\\
hep-th/9411057\\ \end{flushleft}}}
\draft
\tighten
\title{ The Vacuum Structure and Spectrum of\\
$N=2$ Supersymmetric $SU(n)$ Gauge Theory}
\author{Philip C. Argyres\cite{emaila} and Alon E. Faraggi\cite{emailf}}
\address{School of Natural Sciences, Institute for Advanced Study,
Princeton, NJ 08540}
\date{\today}
\maketitle
\begin{abstract}
We present an exact description of the metric on the moduli space
of vacua and the spectrum of massive states for four dimensional
$N=2$ supersymmetric $SU(n)$ gauge theories.  The moduli space of
quantum vacua is identified with the moduli space of a special set of
genus $n-1$ hyperelliptic Riemann surfaces.
\end{abstract}
\narrowtext

Recently Seiberg and Witten \cite{SWI} obtained exact expressions
for the metric on moduli space and dyon spectrum of $N=2$ supersymmetric
$SU(2)$ gauge theory using a version of Olive-Montonen duality \cite{OM}.
In this Letter we use this approach to obtain similar information for
the $N=2$ supersymmetric $SU(n)$ gauge theory with no $N=2$ matter.

The $N=2$ Yang-Mills theory involves a single chiral $N=2$
superfield which, in terms of $N=1$ superfields, decomposes
into a vector multiplet $W_\alpha$ and a chiral
multiplet $\Phi$.  In components, $W_\alpha$ includes the gauge field
strength $F_{\mu\nu}$ as well as the Weyl gaugino, while $\Phi$
includes a Weyl fermion and a complex scalar $\phi$.  All these
fields transform in the adjoint representation of \sun.

The potential for the complex scalar is ${\rm Tr}
[\phi,\phi^\dagger]^2$, implying (at least classically) an
$n-1$-complex dimensional moduli space of flat directions.  Any vev for
$\phi$ can be rotated by a gauge transformation to lie in the Cartan
subalgebra of \sun.  This vev generically breaks $SU(n) \rightarrow
U(1)^{n-1}$.  Denote by $\Phi_i$ and $W_i$ the components of the chiral
superfield $\Phi$ and the vector superfield $W$ in the Cartan
subalgebra with respect to the same basis (so that $\Phi_i$ and $W_i$
are $N=1$ components of the same \uon\ $N=2$ gauge multiplet).  The
running of the couplings of the low-energy \uon's induced by the
symmetry-breaking scales leads to a low-energy effective action
derived from a single holomorphic function ${\cal F}(\Phi_k)$
\cite{S}:
		\be
	\seff = {1\over2\pi}{\rm Im}\left[
	\int d^2\theta\,d^2\overline\theta\,
	\Phi^i_D\,\overline{\Phi_i} +
	{1\over2}\int d^2\theta\, \tau^{ij}\,W_iW_j \right],
		\label{N=2EA}\ee
where, denoting $\partial^i=\partial/\partial \Phi_i$,
		\be
	\Phi^i_D \equiv\partial^i {\cal F},\qquad\qquad
	\tau^{ij} \equiv \partial^i\partial^j {\cal F}.
		\label{DefnAD}\ee
The real and imaginary parts of the lowest component of $\tau^{ij}$ are
the low energy effective theta angles and coupling constants of the
theory respectively.  They are functions of the vevs of the $\phi_i$ fields.

A change in basis of the \uon\ fields corresponding to the
transformation $W_i \rightarrow q^j_i W_j$ by an arbitrary invertible
matrix $\bf q$, could be absorbed in a redefinition $\tau^{ij}
\rightarrow (q^{-1})^i_k \tau^{k\ell} (q^{-1})^j_\ell$ of the effective
couplings.  This ambiguity can be partially fixed by demanding that the
$W_i$ are normalized so that the charges of fields in the fundamental
of \sun\ form a unit cubic lattice so that the allowed set of electric
charges $n^i_e$ are all the integers.  Then the transformations $\bf q$
are restricted to be integer matrices with determinants $\pm1$.
Denoting the magnetic charges of any monopoles or dyons by $2\pi
n_{m,i}$, the Dirac quantization condition requires the $n_{m,i}$ to
lie in the dual lattice to that of the electric charges, implying that
the $n_{m,i}$ are also integers.

The low-energy effective action (\ref{N=2EA}) is left invariant by an
\sptn\ group of duality transformations.  The action of the duality
group on the fields is realized as follows \cite{SWI}.  Define the
$(2n-2)$--component vectors $\t{\bf \Phi}=(\Phi_D^i,\Phi_i)$ and
$\t{\bf W}=(W_D^i,W_i)$, where $W_D^i$ are the dual \uon\ field
strengths.  Then a $(2n-2)\times(2n-2)$ matrix ${\bf M}\in
Sp(2n-2,{\bZ})$ acts as ${\bf \Phi} \rightarrow {\bf M}\cdot{\bf
\Phi}$, ${\bf W} \rightarrow {\bf M}\cdot{\bf W}$, and
$\mbox{\boldmath$\tau$} \rightarrow ({\bf A}\cdot
\mbox{\boldmath$\tau$} + {\bf B}) ({\bf C}\cdot \mbox{\boldmath$\tau$}
+ {\bf D})^{-1}$ where $\bf M={A\,B\choose C\,D}$ and
$\mbox{\boldmath$\tau$}$ denotes the matrix $\tau^{ij}$ of effective
couplings.  With this action one can show \cite{Tata} that any \sptn\
duality transformation can be generated by a change of basis of the
\uon\ generators, the symmetry under discrete shifts in the theta
angles $\tau^{ij} \rightarrow \tau^{ij} + 1$, and the $\tau^{ij}
\rightarrow -(\tau^{ij})^{-1}$ electric-magnetic duality
transformation.

Due to the structure of the $N=2$ supersymmetry algebra \cite{WO}, a
dyon of magnetic and electric charges $\t{\bf n}=(n_{m,i}, n_e^i)$
has a mass saturating the Bogomol'nyi bound \cite{BPS,SWI}
		\be
	M = \sqrt2 \left|\t{\bf a}\cdot{\bf n}\right|\, ,
		\label{BPS}\ee
where $\t{\bf a}=(a_D^i,a_i)$ is the vector of vevs of the scalar
component of the chiral superfield and its dual: $a_i=\langle
\phi_i\rangle$ and $a_D^i=\langle \phi_D^i\rangle$.  This formula is
invariant under the \sptn\ duality transformations since the electric
and magnetic charges transform oppositely to the scalar vevs: ${\bf n}
\rightarrow \t{\bf M}^{-1}\cdot{\bf n}$ if ${\bf a} \rightarrow {\bf
M}\cdot{\bf a}$.

As discussed in Ref.~\cite{SWI}, the combination of the requirements
of analyticity of the superpotential ${\cal F}$ and positivity of the
K\"ahler metric Im{\boldmath$\tau$}, together with the form of the
superpotential at weak coupling, imply that there must be singularities
in the moduli space around which the theory has non-trivial monodromies
lying in \sptn.  Since there is a region of the $SU(n)$ moduli space
where $SU(n)$ is broken at a large scale down to $SU(n-1)$, if follows
that at sufficiently weak coupling a copy of $SU(n-1)$ moduli space
will be embedded in the $SU(n)$ moduli space.  We will essentially use
these facts to find an exact description of the $SU(n)$ moduli space by
induction in $n$.  First, though, we assemble some facts about the
classical $SU(n)$ moduli space.

\paragraph*{Classical Moduli Space.}

The moduli space of the \sun\ theory is most conveniently described in
a basis associated with the $U(n)$ Lie algebra, where the tracelessness
constraint is not imposed.  For this reason we adopt the convention
that upper-case indices $I,J,K,\ldots$ run from 1 to $n$ and lower-case
indices $i,j,k, \dots$ run from 1 to $n-1$.  Use a basis
$\{H^I,E^{IJ}_\pm\, (I>J)\}$ for the generators of the $U(n)$
Lie algebra where the
$n\times n$ matrices $[H^I]_{AB} = \delta_A^I \delta_B^I$ span the
Cartan subalgebra.  Then the \sun\ vector superfield $W = W_I
H^I + W_{IJ}^\pm E_\pm^{IJ}$ will satisfy the tracelessness condition
		\be
	\sum_I W_I = 0.
		\label{TrW}\ee
If we everywhere substitute for $W_n$ in terms of the $W_i$'s using the
tracelessness constraint, we can choose the $W_i$
as a basis of the Cartan subalgebra of \sun.  This basis respects the
requirement imposed in the last section that the charges of fields in the
fundamental of $SU(n)$ generate a unit cubic lattice.

The vev of the complex scalar $\phi$ can always be rotated by a gauge
transformation to lie in the Cartan subalgebra of \sun: $\langle
\phi\rangle = a_I H^I$, where the $a_I$ must also satisfy the
tracelessness constraint
		\be
	\sum_I a_I = 0 .
		\label{Tra}\ee
If we denote the space of independent complex $a_I$'s by ${\cal T}_n =
\{ a_I | \sum_I a_I=0\} \simeq {\bC}^{n-1}$, then the classical
moduli space is ${\cal T}_n$ up to gauge equivalences.  The only
\sun\ elements which act non-trivially on the Cartan subalgebra are the
elements of the Weyl group, isomorphic to the permutation group $S_n$,
which acts by permuting the $a_I$'s.  Thus, the classical moduli space
of the \sun\ theory is ${\cal M}_n = {\cal T}_n/S_n$.

The Higgs mechanism gives the $W^\pm_{IJ}$ bosons masses proportional
to $|a_I-a_J|$.  The Weyl group $S_n$ does not act freely on ${\cal
T}_n$: a submanifold of partial symmetry-breaking to $SU(m)$ is fixed
by $S_m \subset S_n$, since $m$ of the $a_I$'s are equal there.
Classically ${\cal M}_n$ has singularities along these submanifolds
since extra $W_{IJ}^\pm$ bosons become massless there.  Since the
theory is strongly coupled in the vicinity of these submanifolds, one
expects that quantum mechanically the classical moduli space given
above is modified in these regions.

A global U$(1)_{\cal R}$ symmetry of the \sun\ theory is broken down to
${\bZ}_{4n}$ by anomalies.  Since the scalar field $\Phi$ has charge 2
under this symmetry, only a ${\bZ}_{2n}$ acts non-trivially on ${\cal
T}_n$, generated by multiplication of the $a_I$'s by an overall phase
${\rm exp}(i\pi/n)$.  In general the action of the Weyl group $S_n$ and
the global ${\bZ}_{2n}$ do not overlap, except on special curves in
${\cal T}_n$.  Thus, generically, the ${\bZ}_{2n}$ symmetry acts
transitively on the moduli space ${\cal M}_n$.  An exception to this
rule is for $SU(2)$ where a ${\bZ}_2$ of the global ${\bZ}_4$
symmetry coincides everywhere with the Weyl group $S_2\simeq {\bZ}_2$.

A basis of gauge-invariant coordinates covering ${\cal M}_n$ at weak
coupling are given by $u_\alpha = \langle {\rm Tr}( \phi^\alpha )
\rangle = \sum_I a_I^\alpha$, for $\alpha=2,\ldots,n$.
The ${\bZ}_{2n}$ symmetry acts on these coordinates by $u_\alpha
\rightarrow e^{i\pi \alpha/n} u_\alpha$.  A more convenient set of
gauge-invariant coordinates is given classically by the elementary
symmetric polynomials in the $a_I$'s
		\be
	s_\alpha \equiv (-)^\alpha \sum_{I_1<\cdots<I_\alpha}
	a_{I_1}\cdots a_{I_\alpha}, \qquad\alpha=1,\ldots,n.
		\label{symfnc}\ee
These symmetric coordinates can be expressed as polynomials in terms of
the $u_\alpha$'s (thus defining them quantum mechanically).  These
polynomials are generated by Newton's formula
		\be
	r s_r + \sum_{\alpha=0}^r s_{r-\alpha} u_\alpha=0,
	\qquad r=1,2,3,\ldots
		\label{Newt}\ee
where $s_0\equiv 1$, $u_0\equiv 0$,and $s_1=u_1=0$ by the tracelessness
constraint.

\paragraph*{The $SU(n)$ Curve.}

The effective couplings {\boldmath$\tau$} transform under
\sptn\ and Im{\boldmath$\tau$} must be positive definite for
the theory to be unitary.  The period matrix of a genus $n-1$
Riemann surface has precisely these properties, so it is natural
to guess that the moduli space of the $SU(n)$ theory be identified
with the moduli space of the Riemann surface.  Indeed, the solution of
the $SU(2)$ case is of just this form \cite{SWI}.  However, for
$n>2$, the dimension of the moduli space of Riemann surfaces of
genus $n-1$ is too large, so the $SU(n)$ theory must correspond only to
special Riemann surfaces. A relatively simple set of Riemann surfaces are
the hyperelliptic ones \cite{FK}, described by the complex curve
                \be
        y^2 = \prod_{\ell=1}^{2n}(x-e_\ell) ,
                \label{hyper}\ee
which is the double-sheeted cover of the Riemann sphere branched at $2n$
points $e_\ell$.  The $SU(n)$ curve should also have a ${\bZ}_{2n}$ symmetry,
reflecting the $U(1)_{\cal R}$ symmetry broken by instantons in the
$SU(n)$ theory.  This symmetry fits naturally with the hyperelliptic
surfaces if we assign ${\cal R}$-charge 1 to $x$ and $n$ to $y$.

We now assume, following \cite{SWII}, that the coefficients of the
polynomial in $x$ defining the $SU(n)$ curve are themselves polynomials
in the gauge-invariant coordinates $s_\alpha$ (or $u_\alpha$) and
$\Lambda_n^{2n}$, where $\Lambda_n$ is the renormalization scale of
the $SU(n)$ theory.  The power of $\Lambda_n^{2n}$ ensures that it has
the quantum numbers of a one-instanton amplitude.

In the weak coupling limit there are non-trivial monodromies around
the regions of moduli space where extra gauge symmetries are restored.
These regions lie around the submanifolds where
a pair or more of the $a_I$ take the same values.  So, as $\Lambda_n
\rightarrow0$, the $SU(n)$ curve should be singular along these
submanifolds.  A curve is singular whenever a pairs or more of its branch
points $e_\ell$ coincide.   A polynomial in $x$ which has the required
property is $F(x) = \prod_{I=1}^n (x-a_I)$.
As we will shortly see, there is also a monodromy of the $SU(n)$
theory at weak coupling which does not correspond to any classical
singularity of the moduli space.  Thus, in the weak coupling limit
the $SU(n)$ curve should be singular for {\it all} values of the
$a_I$'s.  This can be achieved by simply squaring the polynomial
$F(x)$, so that all its zeros are doubled.  Also, it then has the
right degree in $x$ to desribe a hyperelliptic curve as in (\ref{hyper}).
There is then only one way to add in the instanton contributions
(terms dependent on $\Lambda_n$) consistent with our assignment of
the $\cal R$-charges: $y^2=F^2(x)-\Lambda_n^{2n}$.  The coefficient
of $\Lambda^{2n}_n$ is arbitrary as it reflects a choice of
renormalization group prescription.

It is now easy to extend this curve to strong coupling in $SU(n)$.
The coefficients of the polynomial $F(x)$ are precisely the
elementary symmetric functions $s_\alpha$ of the $a_I$'s (\ref{symfnc}),
which are defined away from weak coupling by Eq.~(\ref{Newt}).
We make the assumption that the $s_\alpha$ remain good global coordinates
on the $SU(n)$ moduli space even at strong coupling.  Then the proposed
$SU(n)$ curve is
	\be
	y^2 = \left( \sum_{\alpha=0}^n s_\alpha x^{n-\alpha} \right)^2
	-\Lambda^{2n}_n.
		\label{curve}\ee

The remainder of this Letter describes various consistency
checks on this proposed curve.  For brevity's sake, we confine
ourselves to checking properties that depend only on the conjugacy
class of the monodromies in \sptn.  A more detailed exposition involving
explicit choices of bases will be given elsewhere \cite{AF}.

\paragraph*{Weak Coupling Monodromies.}

The first check we perform is to show that (\ref{curve}) has all the
right monodromies at weak coupling.  We constructed it only by
demanding that it have singularities at the right places, so computing
the monodromies around those singularities is an independent check.

Note that in the limit where $SU(n)$ is strongly broken down to
$SU(n-1)$, {\it e.g.}\ $a_i \sim a$ and $a_n \sim (1-n)a$ where $|a| >>
\Lambda_n$, then shifting $x$ to $x+a$ in (\ref{curve}) will send two
of the branch points to $\sim -na$ in the $x$ plane while leaving the
rest clustered around the origin.  From the usual renormalization group
matching $\Lambda^{2n}_n \sim a^2 \Lambda_{n-1}^{2(n-1)}$, so taking
the limit $a\rightarrow\infty$ while leaving $\Lambda_{n-1}$ fixed sends
the two branch points at $-na$ to infinity, and rescaling $y$ by
$(x+na)^{-1}$, we recover the curve (\ref{curve}) again, but now for
$SU(n-1)$ instead of $SU(n)$.  Thus the $SU(n)$ curve at weak coupling
automatically contains all $SU(n-1)$ monodromies.  This fact allows us
to proceed by induction in $n$.

First consider the $SU(2)$ curve $y^2=(x^2-\frac{1}{2}u)^2-\Lambda^4$
(where we have used $-2s_2 = u_2 \equiv u$).  This can easily be shown
to be equivalent to the $SU(2)$ curve found in \cite{SWII}, $\tilde y^2
= \tilde x(\tilde x^2+2u\tilde x+\Lambda^4)$, by a fractional linear
transformation on the $\tilde x$ variable.  The point is simply that
the automorphisms of the Riemann sphere allow us to fix three of the
branch points arbitrarily by an $SL(2,\bC)$ transformation.  The
$SU(2)$ curve of Ref.~\cite{SWII} has branch points fixed at 0 and
infinity, whereas the curve (\ref{curve}) does not.

Next consider the $SU(3)$ curve.  We know that along an $SU(2)$ direction
at weak coupling it
degenerates to the $SU(2)$ curve, and so gives the correct monodromies.
However, as mentioned above, the $SU(3)$ curve has another singularity
at weak coupling corresponding to the limit where all the $a_I$'s scale
together by some large factor (or, equivalently, where the $a_I$'s are
held fixed at some generic values and $\Lambda_n\rightarrow0$).  If the
special $SU(3)$ monodromy around this singularity agrees with the
answer calculated from perturbation theory, then all the weak coupling
monodromies of $SU(3)$ will have been checked, and the induction can
proceed to $SU(4)$, {\it etc}.  So, in general, we will need to compute
just one special monodromy for each $SU(n)$ curve.

We are free to pick a convenient curve along which to measure this
monodromy.  Since the special monodromies are not associated with any
coincidences of the $a_I$'s, let us look in a direction in moduli space
along which the $a_I$'s are maximally separated: $a_I = \omega^I a$
where $\omega \equiv e^{2\pi i/n}$.  This is the direction along which
classically all the $s_\alpha$'s except $s_n$ vanish identically.  The
monodromy in question is obtained upon traversing a large circle
at weak coupling in the $s_n$ complex plane.  In this plane the $SU(n)$
curve (\ref{curve}) factorizes for $|s_n| >> \Lambda_n^n$ as
		\be
	y^2 = \prod_{J=1}^n
	\left(x-\omega^J s_n^{1/n} [1+ s_n^{-1}\Lambda_n^n]\right)
	\left(x-\omega^J s_n^{1/n} [1- s_n^{-1}\Lambda_n^n]\right) .
		\ee
The branch points are arranged in $n$ pairs with a pair at each $n$th
root of unity times $s_n^{1/n}$.  As $s_n \rightarrow e^{2\pi i} s_n$,
these pairs are rotated into one another in a counter-clockwise sense,
and each pair also revolves once about its common center in a clockwise
sense.

\iffigs
\begin{figure}[hbtp]
\begin{center}
\leavevmode
\epsfxsize=5cm\epsfbox{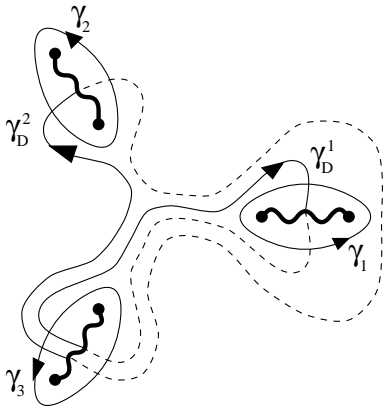}
\end{center}
\caption{Contours for a basis of cycles for the $SU(3)$ curve.
The thick wavy lines represent the cuts, solid contours are on the
first sheet, and dotted ones are on the second.}
\end{figure}
\fi
Choose cuts and a standard basis for the independent cycles on
the $SU(n)$ surface as shown (for $SU(3)$) in Fig.~1.  Thus, $\gamma_1$
and $\gamma_2$ are independent non-intersecting cycles, similarly for
$\gamma_D^i$, and their intersection form is $(\gamma^i_D, \gamma_j) =
\delta^i_j$.  Note that $\gamma_3$ is not independent of the
$\gamma_i$'s:  a simple contour deformation shows that $\sum_I
\gamma_I=0$.  The generalization to the $SU(n)$ curve should be clear.
As $s_n\rightarrow e^{2\pi i}s_n$ the $\gamma_I$ are simply dragged
around the circle so that $\gamma_i \rightarrow \gamma_{i+1} \equiv
P^j_i \gamma_j$, where $P^j_i = \delta^j_{i+1} - \delta^n_{i+1}$ is an
$(n-1)\times(n-1)$ matrix representation of the $\pi=(1\ldots n)$
permutation.

The monodromies of the $\gamma^i_D$ cycles can be determined as
follows.  {}From the monodromies of the $\gamma_i$'s and the defining
properties of symplectic matrices, it follows that the monodromy
{\boldmath$\gamma$}$\rightarrow\bf M\cdot${\boldmath$\gamma$} in
\sptn\ of $\t\mbox{\boldmath$\gamma$}=(\gamma^i_D,\gamma_i)$ can be
written in the block form
		\be
	{\bf M} = \pmatrix{\bf1&\bf N\cr\bf0&\bf1\cr}
	\pmatrix{\t{\bf P}^{-1}&\bf0\cr\bf0&\bf P\cr}
		\label{block}\ee
where $\bf P$ is the permutation matrix found above, and
$\bf N$ is some symmetric matrix which we wish to determine.  Now,
if ${\bf NP}=\t{\bf P}^{-1}\bf N$, so that the two matrices in
Eq.~(\ref{block}) commute, then ${\bf M}^n={{\bf1}\,n{\bf N}\choose
\bf0\,\,\,1\,\,}$ since ${\bf P}^n=1$.  But ${\bf M}^n$ is easy to
compute: as
$s_n \rightarrow e^{2\pi in}s_n$, the $\gamma_i$ cycles are simply
dragged back to themselves and similarly for the $\gamma^i_D$ cycles except
that their ends get wound $n$ times (in a clockwise sense) around each cut
that they pass through.  As illustrated in Fig.~2, each such winding
can be deformed to give two of the associated $\gamma_i$'s.  Keeping
track of the signs, one finds $\gamma^i_D \rightarrow \gamma^i_D -
2n\gamma_i + 2n\gamma_n \equiv \delta^i_j \gamma_D^j + n
N^{ij}\gamma_j$, where
		\be
	N^{ij} = -2(\delta^{ij} +1) .
		\label{monod}\ee
Since (\ref{monod}) satisfies ${\bf NP}=\t{\bf P}^{-1}\bf N$, it
follows that it is, in fact, the matrix $\bf N$ of Eq.~(\ref{block}).
\iffigs
\begin{figure}[hbtp]
\begin{center}
\leavevmode
\epsfxsize=6cm\epsfbox{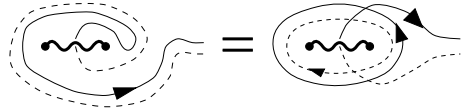}
\end{center}
\caption{Contour unwinding after a long day.}
\end{figure}
\fi

\paragraph*{The Special Monodromies in Perturbation Theory.}

Since pure $N=2$ \sun\ gauge theory is asymptotically free, there is a
weak coupling region where perturbation theory is reliable when
\sun\ is completely broken at a high enough scale so that all the
$|a_I-a_J|>>\Lambda_n$.  We calculate in perturbation theory the
leading behavior of the couplings of the low energy effective action
for the massless $U(1)^{n-1} \subset SU(n)$ gauge bosons.  Denote the
effective coupling of the $W_I$ with $W_J$ fields by $\tilde\tau^{IJ}$,
so the effective $N=1$ gauge action is $\seff\sim\int\tilde\tau^{IJ}W_I
W_J$.  The one-loop result for the running of the couplings is
$\tilde\tau^{IJ} = (i/2\pi)[\delta^{IJ} \sum_K \ln(a_{IK}a_{KI})
-\ln(a_{IJ}a_{JI})]$ where $a_{IJ} \equiv a_I - a_J$.  The
tracelessness constraint (\ref{TrW}) implies $\tau^{ij} =
\tilde\tau^{ij} -\tilde\tau^{in} -\tilde\tau^{nj} +\tilde\tau^{nn}$, or
		\be
	\tau^{ij} = {i\over2\pi}\left\{
	\delta^{ij} \sum_k \ln(a_{ik}a_{ki}) +\delta^{ij}\ln(a_{in}a_{ni})
	-\ln(a_{ij}a_{ji})
	+\sum_k (\delta^{ik}+\delta^{jk}+1) \ln(a_{kn}a_{nk}) \right\} .
		\label{tauij}\ee
{}From Eq.~(\ref{DefnAD}) it follows that
		\be
	a^i_D = \tau^{ij} a_j .
		\label{tad}\ee
A possible constant term in the $a^i_D$ can be shown \cite{SWI} to be
zero by matching to the full \sun\ theory.

In order to compute the monodromies in the $a_D^i$ along a closed
path in ${\cal M}_n$ at weak coupling, we must first lift the path to
a path in ${\cal T}_n$. Since ${\cal M}_n={\cal T}_n/S_n$ is formed by
identifying points in ${\cal T}_n$ which differ by a permutation of
their coordinates $a_I$, in general there will be a non-trivial
monodromy along any path in ${\cal T}_n$ which connects a point with
its image under the action of a non-trivial permutation $\pi\in S_n$.
With one exception, the different possible choices of permutation $\pi$
reflect the pattern of symmetry-breaking of \sun\ at high energies.
For example, the monodromy associated to $\pi=(23\ldots n)$ winds around
a region of moduli space where \sun$\rightarrow SU(n-1)$ at high
energies.  The exception is the monodromy associated to the conjugacy
class of cyclic permutations of all $n$ elements, $\pi=(1\ldots n)$,
which does not correspond to any special symmetry breaking pattern.
This is the monodromy special to $SU(n)$.

As in the computation from the curve, we choose the path realizing the
special monodromy to be $a_I(t)=\omega^{I+t} a$ for $0\leq t\leq 1$,
where $|a|$ is some large scale and $\omega=e^{2\pi i/n}$.
This path precisely traverses a large circle in the
$s_n$ complex plane.  The monodromy of the $a_i$'s along this path is
clearly $a_i \rightarrow P^j_i a_j$, where $\bf P$ is the same
permutation found above from the curve.  The logarithms in
Eq.~(\ref{tauij}) contribute a shift to the $\tau^{ij}$ monodromy,
$\tau^{ij} \rightarrow \tau^{ij} + N^{ij}$, where $\bf N$ is easily
computed to be equal to the $N^{ij}$ of Eq.~(\ref{monod}).  The
$a_D^i$'s then transform as $a_D^i \rightarrow \tau^{ij} P^k_j a_k +
N^{ij} P^k_j a_k$ from (\ref{tad}).  Now, either from the defining
properties of symplectic matrices, or from the fact that the effective
action is completely symmetric among all the low energy $U(1)$'s in the
$s_n$ plane (since $SU(n) \rightarrow U(1)^{n-1}$ at a single scale),
it follows that {\boldmath$\tau$}${\bf P} = \t{\bf
P}^{-1}${\boldmath$\tau$}, and so the monodromy of the scalar vevs
$\t{\bf a}=(a_D^i,a_i)$ indeed agrees with the monodromy (\ref{block})
computed from the $SU(n)$ curve.  This completes our check that the
monodromies of the curve (\ref{curve}) agree with all the monodromies
of the $SU(n)$ theory at weak coupling.

\paragraph*{Metric on Moduli Space and Dyon Spectrum.}

The identification of the metric and spectrum---that is to say,
$a_i$ and $a_D^i$ as functions of the moduli $s_\alpha$---closely
parallels the discussion of Ref.~\cite{SWI}.  Choosing a basis
of cycles $(\gamma_D^i,\gamma_i)$ of the $SU(n)$ curve with the
canonical intersection form $(\gamma^i_D,\gamma_j)=\delta^i_j$,
we identify $a_i$ and $a_D^i$ as sections of a flat \sptn
bundle over moduli space given by
		\be
	a_i = \oint_{\gamma_i} \lambda ,\qquad
	a^i_D = \oint_{\gamma_D^i} \lambda ,
		\label{periods}\ee
where $\lambda$ is some meromorphic one form on the
curve with no residues.  There is a $2n-2$ dimensional space
of such forms spanned by the $n-1$ holomorphic one forms
$(x^{i-1}/y) dx$, and the $n-1$ meromorphic
one forms $x^n \lambda_i$.  The one-form $\lambda$
defining our solution can be written as a linear combination
of these basis one-forms (with coefficients that can depend
on the $s_\alpha$ and $\Lambda_n$) up to a possible total
derivative.

Since the period matrix of the Riemann surface defined by the
$SU(n)$ curve has a positive definite imaginary part, transforms
in the same way as $\tau^{ij}$ under \sptn, and has the same
monodromies as $\tau^{ij}$ does, it follows that they should
be identified.  Now, the period matrix, or $\tau^{ij}$, is
defined by $\sum_j \tau^{ij} (\oint_{\gamma_j}\lambda_k)
= \oint_{\gamma_D^i} \lambda_k$.  Since also
$\tau^{ij} (\partial a_j/\partial s_\alpha) = (\partial a_D^i
/ \partial s_\alpha)$, by (\ref{DefnAD}), it is natural to guess
that
		\be
	{\partial a_i\over\partial s_\alpha} =
	\oint_{\gamma_i}\lambda_\alpha , \qquad
	{\partial a_D^i\over\partial s_\alpha} =
	\oint_{\gamma_D^i}\lambda_\alpha ,
		\label{derivs}\ee
where the $\lambda_\alpha$ are some as yet
undetermined basis of holomorphic one forms.
Eqs.~(\ref{periods}) and (\ref{derivs}) imply a set of differential
equations for $\lambda$.  In the $SU(2)$ case they can be easily
solved to find $\lambda \propto 2x^2(dx)/y$,  since $d\lambda/ds_2
=-(dx)/y +d(x/y)$.
The generalization to $SU(n)$ is\footnote{Special
thanks to R. Plesser who derived this formula.}
	\be
	\lambda \propto\left( \sum_{\alpha=0}^n (n-\alpha) s_\alpha
	x^{n-\alpha}\right) {dx\over y},
	\ee
since $\partial\lambda/\partial s_\alpha = -x^{n-\alpha}(dx)/y +
d(x^{n+1-\alpha}/y)$.  The overall constant
normalization of $\lambda$ can be determined only by making a
choice of basis cycles and matching to perturbation theory.

\paragraph*{Strong Coupling Monodromies.}

The singularities of the curve (\ref{curve}) occur along submanifolds
of the moduli space where a pair
or more of the branch points coincide.  As we have argued above, these
submanifolds all lie at strong coupling.  However, physically,
singularities in the moduli space are expected to occur where a dyon in
the spectrum becomes massless.
The renormalization group flow of the low-energy \uon's to weak
coupling at small scales is cut off at the mass of the lightest charged
particle in the spectrum.  But at those points in moduli space where a
dyon becomes massless, the \uon's that couple to them flow to zero
coupling, and are well-described by perturbation theory.
Thus, there will be a dual description of the physics near the
singular submanifolds which is weakly coupled, and so can be
used to check these limits of the curve (\ref{curve}) as well.

Consider the case where $m$ dyons become massless at a point $P$ in
${\cal M}_n$.  The low energy theory is by definition local, so all $m$
massless dyons must be mutually local.  This implies their charge
vectors ${\bf n}^a$ are symplectically orthogonal: $\t{\bf n}^a \cdot
{\bf I} \cdot {\bf n}^b=0$ for all $a,b=1,\ldots,m$, where $\bf I$ is
the symplectic form ${\bf0\,\,\,1\choose-1\,0}$.  This can only be
satisfied for $m\leq n-1$ linearly independent vectors since there
exists a symplectic transformation to dual fields where each dyon is
described as an electron charged with respect to only one dual low
energy $U(1)$.  In this dual description the physics near the point $P$
is weakly coupled, since $m$ independent electrons are becoming
massless there.

The above symplectic transformation also specifies the dual
scalar vevs $\tilde a_a$ which are good coordinates on moduli space
near $P$ since, by (\ref{BPS}), as $P$ is approached, $\tilde
a_a\rightarrow 0$.  This means that locally in moduli space, a single
dyon, say the one with dual electric charge $\tilde n^1_e$, becomes
massless along a hypersurface of complex co-dimension 1, given by the
solution to the (complex) equation $\tilde a_1=0$.  Two dyons become
massless at the intersection of two such surfaces, which is locally
described as a submanifold of ${\cal M}_n$ of complex codimension 2,
and so forth.  The maximum number $n-1$ of dyons becoming massless at
once will generically occur at an isolated point in moduli space.  Note
that if $m<n-1$, then $n-m-1$ of the \uon's may still be strongly
coupled, and cannot be reliably calculated using perturbation theory.

Along these hypersurfaces the effective action is singular, and so can
lead to nontrivial monodromies for paths looping around them.  The
one-loop effective couplings near $P$ are $\tilde\tau^{ij}=(-i/2\pi)
\delta^{ij} (\tilde n^i_e)^2 \ln(\tilde n^i_e \tilde a_i)$, where
$\tilde n^i_e$ denotes the charge of the $i$th electron.  It is
straightforward to compute the monodromy ${\bf M}_i$ of a path
$\gamma_i$ winding around the $\tilde a_i=0$ hypersurface to be
		\be
	{\bf M}_i = \pmatrix{ {\bf 1}&(\tilde
	n^i_e)^2{\bf e}_{ii}\cr {\bf 0}&{\bf 1}\cr},
		\label{LEMon1}\ee
where ${\bf e}_{ii}$ is an $(n-1)\times(n-1)$ matrix of zeros except
for a 1 in the $i$th position along the diagonal.
A strong coupling test of the curve (\ref{curve}) is that its monodromies
around intersecting singular submanifolds all be conjugate to the
above ${\bf M}_i$ monodromies corresponding to mutually local dyons.

This test for the $SU(2)$ curve is trivially satisfied since the
only singular submanifolds are the two isolated points found in
Ref.~\cite{SWI}.  They each are conjugate to the monodromy (\ref{LEMon1})
with $\tilde n_e = 1$, corresponding to the conjugacy class associated
with the classically stable spectrum of $SU(2)$ dyons.

For the $SU(3)$ curve we first need to identify the singular
submanifolds.  They are given by the vanishing of
the discriminant $\Delta$ of the polynomial $(x^3+s_2x+s_3)^2-\Lambda_3^6$
defining the $SU(3)$ curve.  It is convenient to rescale our
coordinates on moduli space to $\sigma_3 = \Lambda_3^{-3}\, s_3$
and $\sigma_2 = 2^{2/3}3^{-1}\Lambda_3^{-2}\, s_2$.  Then the
discriminant becomes $\Delta(\sigma_2,\sigma_3)=(\sigma_2^3 +
\sigma_3^2)^2 +2(\sigma_2^3 -\sigma_3^2) +1$.  Possible intersection
points of the singular submanifold $\Delta=0$ are at {\it its} singular
points where
$\partial\Delta/\partial\sigma_i = 0$.  There are five such points:
the $\bZ_3$-symmetric triplet of points $\sigma_2^3=-1$ and $\sigma_3=0$,
and the $\bZ_2$ doublet $\sigma_2=0$ and $\sigma_3^2=1$.  The triplet
corresponds to a true intersection point since there $|\partial^2\Delta
/ \partial\sigma_i\partial\sigma_j|\neq 0$.  The $\bZ_2$ points, however,
are not intersection points: in terms of local coordinates $\delta\sigma_i$
vanishing at one of the $\bZ_2$ points, the singular manifold has the
equation $(\delta\sigma_2)^3 = (\delta\sigma_3)^2$.  This describes a
branch point of a single submanifold, instead of the intersection
point of two submanifolds.  Thus, at this point only one dyon
is massless.

We compute the monodromies around the intersecting singular submanifolds
at a $\bZ_3$ point by first expanding the $SU(3)$ curve in local
coordinates around one such point: $s_2\rightarrow -2^{-2/3}3\Lambda_3^2 +
s_2$ and $s_3\rightarrow s_3$, with the new $|s_i|<<\Lambda_3^i$.  Then the
curve approximately factorizes as
		\be
	y^2\sim(x-1+\sqrt{s_2+s_3}) (x-1-\sqrt{s_2+s_3})
	(x+1+\sqrt{s_2-s_3}) (x+1-\sqrt{s_2-s_3}) (x^2-4)
		\ee
where we have rescaled $\Lambda_3 \rightarrow 2^{1/3}$.  Choose
a basis of $\gamma_i$ cycles to encircle the pairs of branch points near
$-1$ and $+1$, and the $\gamma_D^i$'s in the canonical way.  Paths
which encircle the intersecting singular manifolds are simply a
circle in the $s_2+s_3$ complex plane keeping $s_2-s_3$ fixed,
and {\it vice versa}.  The resulting monodromies are then easily
found to be precisely of the form (\ref{LEMon1}) with $\tilde n_e^1 =
\tilde n_e^2 = 1$.  This confirms that there are indeed two different
mutually local dyons becoming massless along the two intersecting
submanifolds at the $\bZ_3$ points.  Furthermore, their charges are
consistent with the semi-classically stable dyon charges in the
$SU(2)$ limit.  This suggests that, as in the $SU(2)$ case, the spectrum
of stable dyon charges remains the semi-classical one all the way
down to these strong-coupling singularities.

As a final check of the $SU(3)$ curve, we note that the $\bZ_3$
intersection points imply the known $N=1$ $SU(3)$ vacuum structure.
Indeed, following the arguments of Ref.~\cite{SWI}, add to the
microscopic $N=2$ theory a coupling $\mu$ to the composite $N=1$
superfield corresponding to $s_2$.  This is a mass term for the
$N=1$ chiral superfield $\Phi$.   Going to the dual (weakly coupled)
description of the physics near a point in the moduli space of the
$SU(n)$ theory where $n-1$ dyons are massless, and using the
non-perturbative nonrenormalization theorem of \cite{Snr},
the non-perturbative form of the effective superpotential
is found to be ${\cal W}= \sum_i \tilde a_i(s_\alpha) m_i \tilde m_i +
\mu s_2$, where $m_i$ and $\tilde m_i$ are the lowest components
of the dyon chiral superfields.  Minimizing the superpotential
subject to the D-term constraints $|m_i|=|\tilde m_i|$ for all $i$
shows that for non-zero $\mu$ the $N=2$ flat directions are lifted
and only the point $\tilde a_i=0$ where all $n-1$ dyons are massless
remains an $N=1$ vacuum.  The three $\bZ_3$ singularity intersection
points of the $SU(3)$ curve found above are just such points, and
happily they correspond to the three $N=1$
$SU(3)$ vacua related by a spontaneously broken $\bZ_3$.

Computing the discriminant and finding all the strong coupling
singularities for the $SU(n)$ curve becomes increasingly difficult
for higher $n$.

\paragraph*{Acknowledgements.}

It is a pleasure to thank P. Berglund, R. Dick, J. March-Russell,
E. Witten, and especially R. Plesser, N. Seiberg, and A. Shapere for many
helpful discussions and comments.  The work of P.C.A. is supported by
NSF grant PHY92-45317 and by the Ambrose Monell Foundation.
The work of A.E.F. is supported by DOE grant DE-FG02-90ER40542.

\paragraph*{Note Added.}  In the course of writing this paper there
appeared preprint {\tt hep-th/9411048} by A. Klemm, {\it et. al.}
which studies the same problem.


\begin{references}
\bibitem[*]{emaila}argyres@guinness.ias.edu.
\bibitem[**]{emailf}faraggi@sns.ias.edu.
\bibitem{SWI}N. Seiberg and E. Witten, {\sl Nucl. Phys.} {\bf B426} (1994)
19; {\tt hep-th/9407087}.
\bibitem{OM}C. Montonen and D. Olive, {\sl Phys. Lett.} {\bf 72B} (1977)
117; P. Goddard, J. Nuyts, and D. Olive, {\sl Nucl. Phys.} {\bf B125}
(1977) 1.
\bibitem{S}N. Seiberg, {\sl Phys. Lett.} {\bf 206B} (1988) 75.
\bibitem{Tata}See for example D. Mumford, {\it Tata Lectures on Theta I},
Birkh\"auser (1983).
\bibitem{WO}E. Witten and D. Olive, {\sl Phys. Lett.} {\bf 78B} (1978) 97.
\bibitem{BPS}M. K. Prasad and C. M. Sommerfield, {\sl Phys. Rev. Lett.}
{\bf 35} (1975) 760; E. B. Bogomol'nyi, {\sl Sov. J. Nucl. Phys.} {\bf
24} (1976) 449.
\bibitem{FK}See for example H. M. Farkas and I. Kra, {\it Riemann
Surfaces}, Springer-Verlag (1980).
\bibitem{SWII}N. Seiberg and E. Witten, {\it Monopoles, Duality, and
Chiral Symmetry Breaking in N=2 Supersymmetric QCD}, preprint
RU-94-60, IASSNS-94/55, {\tt hep-th/9408099}.
\bibitem{AF}P. C. Argyres and A. E. Faraggi, {\it Duality and the
Vacuum Structure of N=2 Supersymmetric Yang-Mills Theory,} preprint
IASSNS-HEP-94/95, to appear.
\bibitem{Snr}N. Seiberg, {\sl Phys. Lett.} {\bf 318B} (1993) 469; {\tt
hep-th/9309335}.
\end{references}
\end{document}

\end